\documentclass[traditabstract]{aa} % for the abstract without structuration 
\usepackage{amsmath}
\usepackage{graphicx}
\usepackage{txfonts}

\usepackage{natbib}
\bibpunct{(}{)}{;}{a}{}{,}
\usepackage{float}
\usepackage{calc}
\usepackage{textcomp}
\usepackage{placeins}
\usepackage{color}
\usepackage{rotating} 
\usepackage{lscape}
\usepackage{url}
\usepackage[colorlinks=true,linkcolor=blue,citecolor=blue]{hyperref}
\usepackage{subfig}
\usepackage[all]{draftcopy}
\usepackage{longtable}
\usepackage{array}
\usepackage{threeparttable}
\usepackage{lscape}
\DeclareTextSymbol{\degre}{OT1}{23}
\newcounter{savedfootnote}

\def \microns{{\,$\mu$m}}
\def \HI{{H{\sc i}}}
\def \HIdef{{H{\sc i}$-def$}}

\begin{document}
   \title{The imprint of rapid star formation quenching on the spectral energy distributions of galaxies.}

\author{L.~Ciesla\inst{1,2,3},
	A.~Boselli\inst{4},
	D.~Elbaz\inst{3},			
	S.~Boissier\inst{4},
	V.~Buat\inst{4},
	V.~Charmandaris\inst{2,1},
	C.~Schreiber\inst{3},
	M.~B\'ethermin\inst{5},
	M.~Baes\inst{6},
	M.~Boquien\inst{7},
	I.~De Looze\inst{8,7,6},
	J.A.~Fern\'andez-Ontiveros\inst{9},
	C.~Pappalardo\inst{10},
	L.~Spinoglio\inst{9},
	S.~Viaene\inst{6}.
}

\institute{	
  University of Crete, Department of Physics, Heraklion 71003, Greece
 \and
 Institute for Astronomy, Astrophysics, Space Applications and Remote Sensing, National Observatory of Athens, GR-15236 Penteli, Greece
  \and
  Laboratoire AIM-Paris-Saclay, CEA/DSM/Irfu - CNRS - Université Paris Diderot, CEA-Saclay, F-91191 Gif-sur-Yvette, France
  \and
  Aix-Marseille  Universit\'e,  CNRS, LAM (Laboratoire d'Astrophysique de Marseille) UMR7326,  13388, Marseille, France	
 \and
 European Southern Observatory, Karl-Schwarzschild-Str. 2, 85748, Garching, Germany
 \and
Sterrenkundig Observatorium, Universiteit Gent, Krijgslaan 281 S9, B-9000 Gent, Belgium
 \and 
 Institute of Astronomy, University of Cambridge, Madingley Road, Cambridge CB3 0HA, UK
 \and
 Department of Physics and Astronomy, University College London, Gower Street, London WC1E 6BT, UK
 \and
 Istituto di Astrofisica e Planetologia Spaziali (INAF-IAPS), Via Fosso del Cavaliere 100, I-00133, Roma, Italy
 \and
 Centro de Astronomia e Astrofísica da Universidade de Lisboa, Observat\'orio Astron\'omico de Lisboa, Tapada da Ajuda, 1349-018, Lisbon, Portugal
}	
 
   \date{Received; accepted}

  \abstract
{
In high density environment, the gas content of galaxies is stripped, leading to a rapid quenching of their star formation activity.
This dramatic environmental effect, not related to typical passive evolution, is generally not taken into account in the star formation histories (SFH) usually assumed to perform spectral energy distribution (SED) fitting of these galaxies, yielding to a poor fit of their stellar emission and, consequently, a biased estimate of the star formation rate (SFR).
In this work, we aim at reproducing this rapid quenching using a truncated delayed SFH that we implemented in the SED fitting code CIGALE.
We show that the ratio between the instantaneous SFR and the SFR just before the quenching ($r_{SFR}$) is well constrained as long as rest frame UV data are available.
This SED modelling is applied to the \textit{Herschel} Reference Survey (HRS) containing isolated galaxies and sources falling in the dense environment of the Virgo cluster.
The latter are \HI-deficient due to ram pressure stripping.
We show that the truncated delayed SFH successfully reproduces their SED while typical SFH assumptions fail.
A good correlation is found between $r_{SFR}$ and \HIdef, the parameter quantifying the gas deficiency of cluster galaxies, meaning that SED fitting results can be used to provide a tentative estimate of the gas deficiency of galaxies for which \HI\ observations are not available.
The HRS galaxies are placed on the SFR-$M_*$ diagram showing that the \HI-deficient sources lie in the quiescent region confirming previous studies.
Using the $r_{SFR}$ parameter, we derive the SFR of these sources before quenching and show that they were previously on the main sequence relation.
We show that the $r_{SFR}$ parameter is also well recovered for deeply obscured high redshift sources, as well as in absence of IR data.
SED fitting is thus a powerful tool to identify galaxies that underwent a rapid star formation quenching.
 
}

  
   \authorrunning{Ciesla et al.}
   \titlerunning{Star formation quenching from SEDs}

   \maketitle

%=================================================================================
\section{\label{intro}Introduction}

At low redshifts, the bimodality of galaxies in the optical-to-ultraviolet (UV) colour \citep[proxy to the star-formation history, e.g.][]{Salim05} versus stellar-mass plane is now established and has been interpreted as evidence for star-formation quenching \citep[e.g.][]{Strateva01,Baldry04,Bell04,Baldry06,GildePaz07}.
However, the origin of this quenching is still open for debate.
External processes linked to the environment have been proposed as one quenching mechanism from the observation that high density environments are occupied mainly by early-type galaxies \citep[e.g.,][]{Dressler80}.
One example of environment-driven quenching is ram pressure stripping, a process expected to take place in the dense environment of clusters.
When a galaxy moves through the hot gas with a significant speed, its gas content is being stripped out \citep[e.g.,][]{GunnGott72,MoriBurket00,Quilis00,Kronberger08,Bekki09,TonnesenBryan09}.
This mechanism is observed in galaxies of nearby clusters such as  the Virgo and the Coma clusters \citep[e.g.,][ and references therein]{BoselliGavazzi14} and it is known that these objects lie on the green valley and are considered, in terms of star formation, intermediates between normal star forming discs and passive early-type galaxies \citep[e.g.,][]{Boselli08,HughesCortese09,CorteseHughes09}.
Recently, \cite{Boselli14} showed that the molecular gas is also affected by ram pressure stripping with the most affected galaxies having, on average, a factor of $\sim$2 less molecular gas than similar objects in the field.
Furthermore, \textit{Herschel} observations have shown that the dust content of \HI-deficient galaxies is also affected by the cluster environment \citep{Cortese10b,Cortese12a}.
Thus, due to ram pressure, both the \HI\ and the molecular gas are stripped out, leading to a quenching of the star formation activity.
Observational evidence showed that this quenching can be rapid as it can take ram pressure about several hundreds of Myr to produce a typical gas deficient cluster galaxy  and less than $\sim$1.5\,Gyr to completely remove the gas \citep{Vollmer04,Vollmer08,Vollmer09,Vollmer12,Boselli06,RoedigerBruggen07,CrowlKenney08,Pappalardo10,Boselli14}.
However, other studies based on simulations and observations suggested that ram pressure could act on longer timescales rather than instantaneously \citep[e.g.,][]{McGee09,Weinmann10,Haines13,Haines15}.
Determining the timescales over which quenching mechanisms are efficient is thus critical. 

The reduction of the star formation rate of a galaxy, rapid or smooth, is translated in its star formation history (SFH).
Thus  the quenching affects the shape of the UV to near-infrared (NIR) emission of its stellar population.
Indeed, it is well established now that the NUV-r color is also an excellent tracer of gas content \citep{Cortese11,Fabello11,Catinella13,Brown15}.
Consequently, fitting the observed UV to NIR emission of galaxies with stellar population models convolved by assumed SFHs allow us to place constrains on the true galaxy SFH.
Such methods have been widely applied in the literature to retrieve key physical properties of galaxies such as their stellar masses and star formation rates (SFR).

To model a galaxy  SED, it is necessary to make some assumptions  about its  star-formation history.   
Galaxies typically form their bulk of stars through secular process with, in addition, variations of the SFR on different timescales, such as star formation bursts \citep[e.g.,][]{Elbaz11,Sparre15}.
Simple functional  forms are usually assumed in the literature, such  as an exponentially  decreasing or increasing  SFR,  two exponential  decreasing  SFRs with  different  e-folding  times, a delayed  SFH, a lognormal SFH, or an instantaneous burst \citep[e.g.,][]{Ilbert13,Pforr12,Lee09,Schaerer13,Gladders13}.  
These simple assumptions reproduce only smooth variations of the SFH, even though it is possible to add one or several burst of star formation.
None of these SFHs though can model a rapid and quite sudden quenching of star formation.

In this work, we use a new SFH analytical function to model a rapid and sudden quenching of star formation in galaxies.
To evaluate this approach, we need a sample of galaxies which have experienced this rapid quenching scenario.
Nearby galaxies of the Virgo cluster are known to undergo ram pressure stripping and are thus good candidates for our study.
Some of these galaxies are included in the \textit{Herschel} Reference Survey \citep{Boselli10a} which benefit from a wealth of ancillary data and contains galaxies from a wide range of environments.
This stripping of the gas, observed in late-type cluster galaxies, can be quantified by the \HI-deficiency. 
This parameter is defined as the logarithmic difference between the average HI mass of a reference sample of isolated galaxies of similar type (from Sa to Scd-Im-BCD) and linear dimension and the \HI\ mass actually observed in individual objects: \HIdef $= \log M$\HI$_{ref} -  \log M$\HI$_{obs}$.
According to \cite{Haynes84}, $\log h^2 M$\HI$_{ref} = c + d \log(h\times diam)^2$, where $c$ and $d$ are weak functions of the Hubble type, $diam$ (in kpc) is the linear diameter of the galaxy \citep[see][]{Gavazzi05,BoselliGavazzi09} and $h = H_0/100$.
In the following, we define as ``gas-rich'' or ``normal'' galaxies with \HIdef$\leq 0.4$, and ``\HI-deficient'' those with \HIdef$> 0.4$, following \cite{Boselli12}.
Although the HRS sample was first built to study the dust properties of galaxies with \textit{Herschel}, the large photometric coverage available for these sources, from UV to submm wavelengths, makes it ideal to test a new SED modeling approach.

The paper is organized as follows.
We present the \textit{Herschel} Reference Survey in Sect.~\ref{hrs} and the SED fitting code CIGALE in Sect.~\ref{cig}.
In Sect.~\ref{trunc}, we present the truncated SFH included into CIGALE and study the constraints that we have on its free parameters.
The results of the SED fitting are presented in Sect.~\ref{sedfit} and are discussed in Sect.~\ref{discussion}.
SED fitting is performed assuming an IMF of \cite{Salpeter55}, but the results of this work are found to be robust against IMF choice.
%=================================================================================

%=========
\section{\label{hrs}The \textit{Herschel} Reference Survey sample}

To examine the impact of rapid quenching on the SED of galaxies, we use the \textit{Herschel} Reference Survey \citep[HRS,][]{Boselli10a}.
The HRS is a combined volume- and flux- limited sample composed of galaxies with a distance between 15 and 25\,Mpc.
The galaxies are then selected according to their K-band magnitude, a reliable proxy for the total stellar mass \citep{Gavazzi96}. 
The sample contains 322\footnote{With respect to the original sample given in \cite{Boselli10a}, the galaxy HRS\,228 is removed from the complete sample because its updated redshift on NED indicates it as a background object.}  galaxies, among which 62 early-type and 260 late-type.
The HRS covers all morphological types, we use the classification presented in \cite{Cortese12b}.

The \HI\ data available for the HRS galaxies are presented in \cite{Boselli10a}.
They are available for 96\% of the late-type galaxies of the HRS.
\HI\ deficiencies of the HRS galaxies have been measured using the calibrations of \cite{BoselliGavazzi09}.
The HRS galaxies have \HIdef\ between -0.65 and 1.69.
The gas rich galaxies will serve as a comparison element to identify possible trends linked to the gas deficiency.
This reference sample benefits from a wealth of ancillary data, photometric from UV to radio \citep[][and from the literature]{Bendo12a,Cortese12b,Ciesla12,Cortese14,Ciesla14} as well as optical spectra \citep{Boselli13,Boselli14a,Boselli15}.

In this work, we select the HRS galaxies defined as late-type (Sa and later types) which are also detected with GALEX in the FUV and NUV, since these observations are mandatory for our study, as we will discuss in Sect.~\ref{cons}.
Our subsample consists of 228 galaxies, 136 normal star-forming and 92 \HI-deficients.
The photometric coverage of this subsample is presented in Table~\ref{filt} and is thoroughly discussed in the associated papers.
Even though a direct consequence of ram pressure stripping is evident through the observed UV to NIR emission, a complete modeling of the galaxy SED, using data from the UV to the sub-mm, is essential to quantify the current stellar emission \cite[e.g.][]{Buat14}.

\begin{table}
	\centering
	\caption{Broad-band filter-set used in this
                  paper. }
	\begin{tabular}{ l l c l }
	 \hline\hline
	Telescope/Camera & Filter Name & $\lambda_{mean}$(\microns) & Ref.\\ 
	\hline
	GALEX & FUV &  0.153 & a \\
				 & NUV & 0.231& a \\
				& U & 0.365 & b\\
				& B & 0.44 & b\\
	SDSS	& g & 0.475 & a\\
				& V & 0.55 & b\\
	SDSS	& r & 0.622 & a\\
	SDSS	& i	& 0.763 & a\\		 
	2MASS &  J & 1.25 & b \\
			 &  H & 1.65 & b \\
			 & Ks & 2.1 &  b \\
	\textit{Spitzer}& IRAC1& 3.6 & c\\
			& IRAC2 & 4.5  & c \\
			& IRAC4 & 8 & d \\
	WISE & 3	& 12  & d \\
			& 4 & 22 &  d\\
			& MIPS1 & 24 & e\\
			& MIPS2 & 70  & e\\
	\textit{Herschel}& PACS green & 100 & f\\
			& PACS red & 160 & f\\
			& PSW & 250 & g\\
			& PMW & 350 & g\\
			& PLW & 500 & g\\
	\hline
	\label{filt}
	\end{tabular}
	\tablefoot{
	\tablefoottext{a}{\cite{Cortese12b}.}
	\tablefoottext{b}{Compilation from the literature, details are provided in \cite{Boselli10a}.}
	\tablefoottext{c}{S$^4$G: \cite{Querejeta15}.}
	\tablefoottext{d}{\cite{Ciesla14}.}
	\tablefoottext{e}{\cite{Bendo12a}.}
	\tablefoottext{f}{\cite{Cortese14}.}
	\tablefoottext{g}{\cite{Ciesla12}.}
	}
\end{table}

%=================================================================================
\section{\label{cig}Modelling and fitting galaxy SEDs with CIGALE}

CIGALE\footnote{The code is publicly available at: \url{http://cigale.lam.fr/}.}  (Code Investigating GALaxy Emission) is  a SED modeling software package  that  has  two functions:  a SED modeling  function and a SED fitting function \citep[][Burgarella et al., in prep; Boquien et al., in prep]{Roehlly14}.   
Even though the philosophy of CIGALE, originally presented in \cite{Noll09}, remains, the code has been rewritten in Python and additions have been made in order to optimize its performance and broaden its scientific applications. 
The  SED modeling  function of CIGALE allows the building of galaxy SEDs from the UV to the radio by assuming a combination of modules which model the star formation history (SFH) of the galaxy, the stellar emission from stellar population models \citep{BruzualCharlot03,Maraston05}, the nebular lines, the attenuation by dust \citep[e.g.,][]{Calzetti00}, the IR emission from dust \citep{DraineLi07,Casey12,Dale14}, the AGN contribution \citep{Fritz06}, and the radio emission.  
CIGALE builds the SEDs taking into account the balance between the energy absorbed by dust and reemitted in the IR.

These modeled SEDs are then integrated into a set of filters to be compared directly to the observations.
For each parameter, a probability distribution function (PDF) analysis is made. 
The output value is the likelihood-weighted mean value of the PDF and the error associated is the likelihood-weighted standard deviation. 
We use CIGALE to derive the physical properties of galaxies such as stellar masses, instantaneous SFRs, dust attenuation, IR luminosities, dust masses, taking into account panchromatic information on the SED.
	
In CIGALE, the assumed SFH can be  handled in two different ways. 
The first is to model it using simple analytic functions (e.g. exponential forms, delayed  SFHs, etc).
The second is to provide more complex SFHs, such as  those provided by  semi-analytical models (SAM) and simulations \citep{Boquien14,Ciesla15}, an approach which we will not use in the present study.

%=================================================================================
\section{\label{trunc}Truncated SFH}

\begin{figure}%[!h] 
  	\includegraphics[width=\columnwidth]{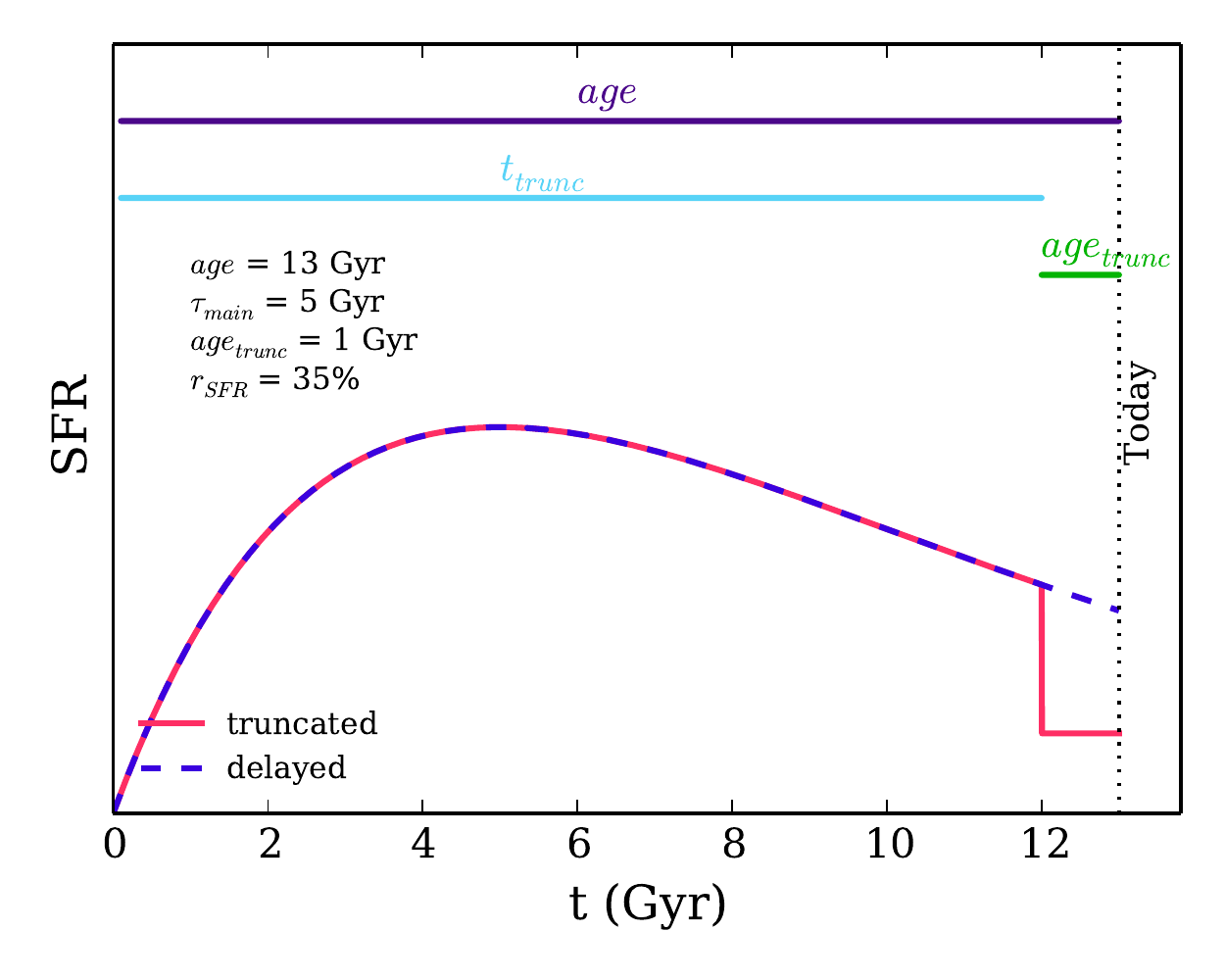}
  	\caption{ \label{sfh_schema} Illustration of the truncated delayed SFH implemented in CIGALE. The purple dashed line represents a normal delayed SFH with $\tau_{main}$=5\,Gyr, without truncation. The red solid line is the truncated delayed SFH.}
\end{figure}

\begin{figure*}%[!h] 
  	\includegraphics[width=\textwidth]{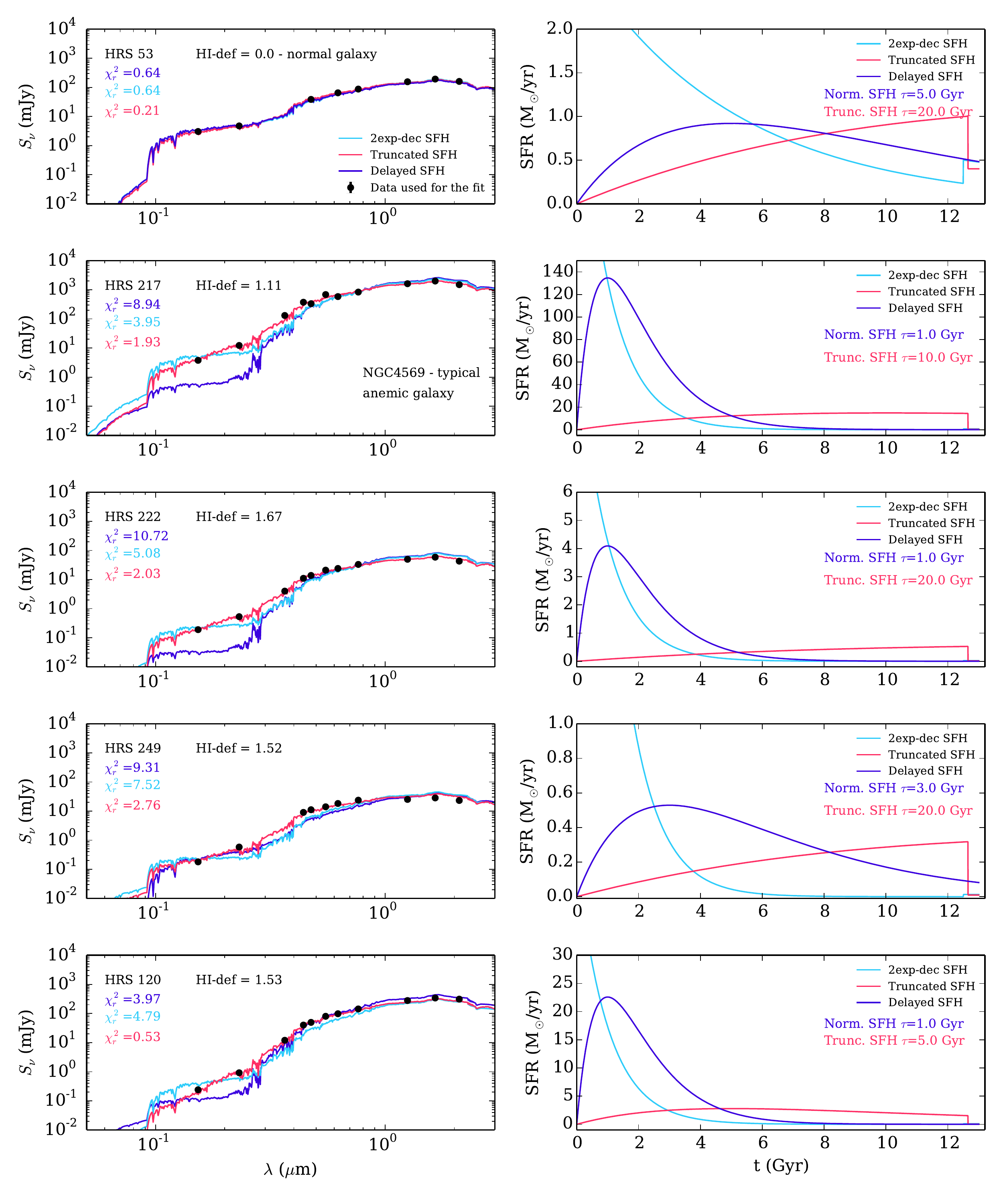}
  	\caption{ \label{sedex} Left column: examples of best fit obtained with CIGALE for one normal and four \HI-deficient HRS galaxies using three different SFHs assumptions: a delayed SFH (navy blue), a 2-exponentially decreasing SFH (cyan) and a truncated delayed SFH (red). Right column: for each galaxy, the output best fit SFH obtained for the three different assumptions made.}
\end{figure*}
	
	%=========
	\subsection{The need for truncated SFH}
	
	Several studies have already shown that the delayed SFH matches observations as well as SFHs obtained through SAM or hydrodynamical simulations \citep[e.g.,][]{Boselli01,Sparre15,Ciesla15}.
	We thus base our work on the delayed SFH which is represented with the following expression:
	\begin{equation}
		SFR(t) \propto t\,exp(-t/\tau_{main}).
	\end{equation}
	\noindent where $t$ is the time ($t=0$ corresponds to the time when the first stars of the galaxy formed) and $\tau_{main}$ is the e-folding time of the stellar population.
	As shown in Fig.~\ref{sfh_schema}, this SFH describes secular star formation but not any star formation burst.
	Stellar masses and SFR obtained through SED fitting using this SFH yield offsets less than 10\%, comparable to the widely spread exponentially decreasing SFHs (with one or two stellar populations),
	furthermore, delayed SFH provides better estimates of the age of a galaxy contrary to exponentially decreasing models which underestimate this parameter \citep{Maraston10,Pforr12,Ciesla15}
		
	Even though the SFHs assumptions commonly used in the literature provide good fits of the stellar emission of normal galaxies, they fail to reproduce the stellar emission of rapidly quenched galaxies such as the \HI-deficient galaxies of the HRS.
	Indeed, Fig.~\ref{sedex} shows the best fits obtained for one normal and four \HI-deficient galaxies using the delayed and, for comparison, the usual 2-exponentially decreasing SFH.
	The dynamical range of the parameters and their sampling are presented in Table~\ref{fitparam}.
	These two models fit well the data of the normal (gas-rich) galaxy (top panel).
	 However, the delayed SFH model struggles to reproduce simultaneously the UV-optical observations linked to the young stellar population and the NIR data linked to the old stellar population.
	 Furthermore, to be able to reproduce the optical-NIR data as well as the very low SFR, the delayed SFH obtained from the best fits have overall a low value of $\tau_{main}$ (Fig.~\ref{sedex}, right column).	A delayed SFH with a low value of $\tau_{main}$ is more characteristic of early-type passive galaxies with the creation of the bulk of the stars at early time and then a smooth decrease of the star formation activity.
	The 2-exponentially decreasing SFH model better reproduces the UV emission but with the consequence of underestimating the optical data.
	We also explore leaving the age of the galaxy as a free parameter and do not find any change in the results shown in Fig.~\ref{sedex}.
	Thus, the usually assumed SFHs fail to reproduce the peculiar UV-NIR SED of rapidly quenched galaxies such as the \HI-deficient objects.

	%=========
	\subsection{\label{implsfh}Implementing a truncated SFH}

	In order to model the SFH of  \HI-deficient galaxies, previous studies \citep[e.g.,][]{Boselli06,Fumagalli11} proposed to use a delayed SFH which, at a given time, would have a strong decrease of the SFR.
	We thus include in CIGALE such a SFH, with the following expression:

	\begin{equation}	
    	SFR(t) \propto
    	\begin{cases}
    	  t \times exp(-t/\tau_{main}), & \text{when}\ t<=t_{trunc} \\
    	  r_{SFR} \times SFR(t=t_{trunc}), & \text{when}\ t>t_{trunc} \\
    	\end{cases}
	\end{equation}	
	
	\noindent where $t_{trunc}$ is the time at which the star formation is quenched, and $r_{SFR}$ is the ratio between SFR$(t>t_{trunc})$ and SFR$(t=t_{trunc})$:
	
	\begin{equation}
	 r_{SFR} = \frac{SFR(t>t_{trunc})}{SFR(t_{trunc})}.
	\end{equation}
	\noindent 	Fig.~\ref{sfh_schema} shows an example of truncated SFH.
	
	The SFH is thus determined through four parameters, the age of the galaxy, the e-folding time of the main stellar population model, $\tau_{main}$, the age of the truncation, $age_{trunc}$, and $r_{SFR}$.
	This model representing  the effect of the cluster environment on galaxy SFH is simple in order to limit the possible degeneracy that could arise from a more complex shape with additional free parameters. 
	We will discuss its validity in Sect.~\ref{discussion}.

\begin{figure}%[!h] 
  		\includegraphics[width=\columnwidth]{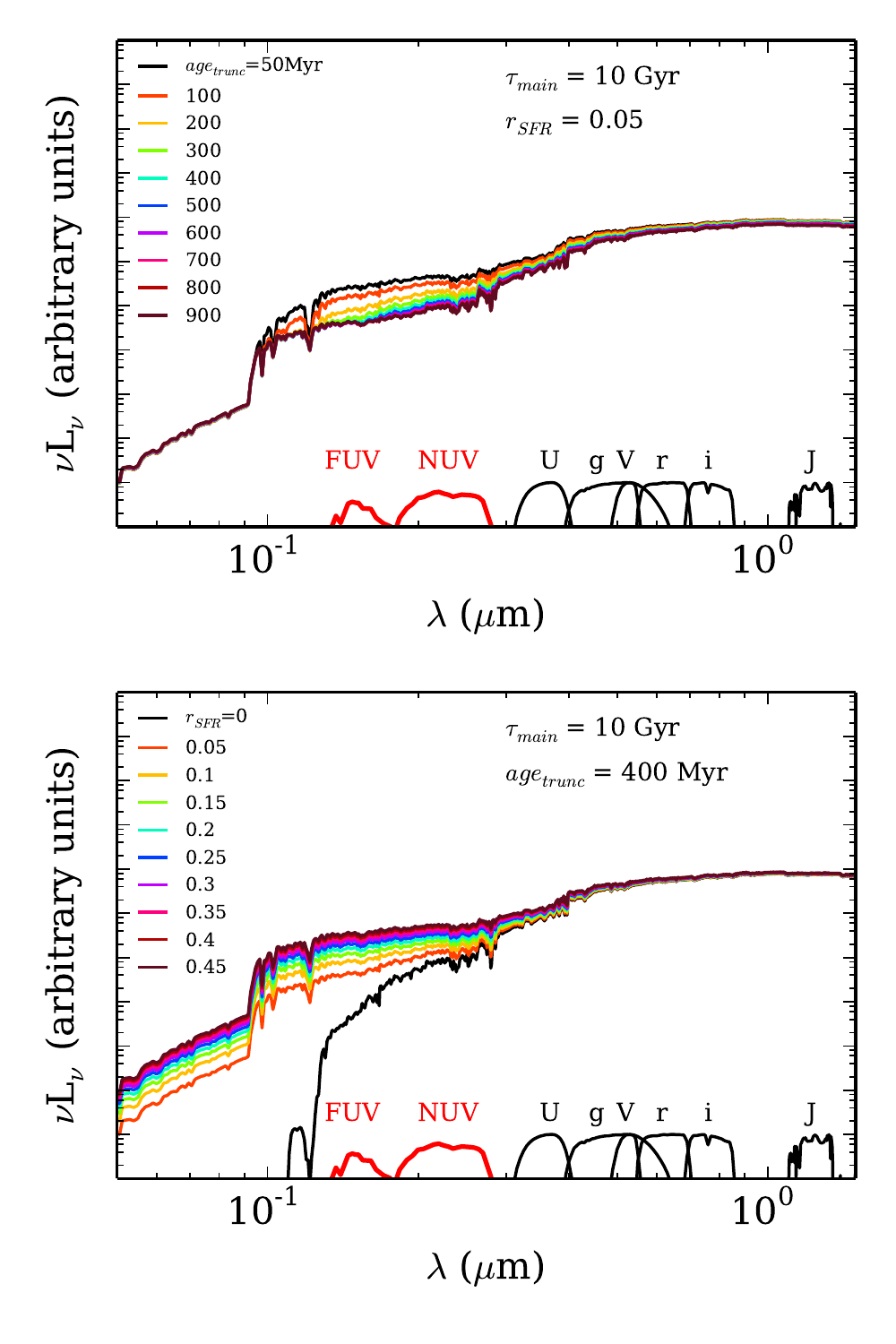}
  		\caption{ \label{models} Impact of the two parameters, $age_{trunc}$ (upper panel) and $r_{SFR}$ (lower panel), on the SED considering a $\tau_{main}$ of 10\,Gyr. SEDs are color coded according to the value of the free parameter. For clarity, we only show in this plots models with $r_{SFR}$ lower than 0.45. At the bottom of each panel, we show the filters of GALEX (red), SDSS, and 2MASS.}
	\end{figure}

	In order to understand the impact of the two parameters handling the truncation, $age_{trunc}$ and $r_{SFR}$ on the shape of the SED, we show in Fig.~\ref{models} modeled UV-to-optical SEDs varying the truncation age (top panel) and the ratio between the SFR after and before the truncation (bottom panel).
	The truncation age, $age_{trunc}$ mainly impacts the SED between 0.1 and 0.5\microns\ with a diminution of the emission in this range when $age_{trunc}$ increases.
	The $r_{SFR}$ parameter impacts the SED at wavelengths shorter than 0.5\microns.
	When the SFR after truncation is null, the emission at $\lambda<0.1$\microns\ drops significantly.
	Very small values of $r_{SFR}$, 0.05 for instance, are sufficient to produce emission at $\lambda<0.1$\microns, that increases with $r_{SFR}$.
	From Fig.~\ref{models}, it is clear that the GALEX filters are mandatory to constrain these two parameters as both FUV and NUV filters probe the spectral range where they impact the SED.

	%=========
	\subsection{\label{cons}Evaluating the SFH parameters with a mock catalogue}
	
	To examine if we can constrain these four parameters from broad band SED fitting with CIGALE as well as the accuracy and precision that we can expect, we build a mock galaxy catalogue.
	A mock galaxy catalogue  consists of theoretical SEDs, here built with CIGALE, for which we know the exact underlying physical parameters (i.e. galaxy age, complete SFH, etc). 
	Using our fitting procedure on this mock catalogue allows us to compare the output results to the parameters used to build the mock catalogue, and evaluate how well we constrain a given parameter.
	We use the modeling function of CIGALE to create a set of galaxy SEDs, following the method presented in \cite{Giovannoli11}.
	The stellar emission is computed convolving the stellar population models of \cite{BruzualCharlot03} with the truncated delayed SFH presented in Sect.~\ref{implsfh}.
	We use the \cite{Calzetti00} law to attenuate the stellar emission while the IR emission is modeled with \cite{Dale14} templates.  
	The parameters used to compute the mock SEDs are presented in Table~\ref{fitparam}, in bold.
	The mock SEDs are integrated into a set of filters corresponding to the observations available for the HRS galaxies (Table~\ref{filt}).
	Following the method presented in \cite{Ciesla15}, we perturb the mock flux densities adding a noise randomly taken from a Gaussian distribution with $\sigma$=0.1 and a photometric error of 15\% is assumed for each flux density.
	
	\begin{table}
		\centering
		\caption{Galaxy parameters used in the SED fitting procedures. The values in bold are used to generate the mock galaxy catalogue of SEDs.}
		\begin{tabular}{l c }
	 	\hline\hline
		Parameter & Value \\ 
		\hline
		\multicolumn{2}{c}{Double exp. decreasing SFH\tablefootmark{a}	}\\  
		\hline
		$age$ (Gyr) & 13	\\
		$\tau_{main}$  (Gyr) & 0.5, 1, 3, 5, 20\\
		$f_{burst}$ 	& 0.001, 0.01, 0.1, 0.99	\\ 
		$age_{burst}$ (Myr) 	& 50, 100, 200, 500\\ 
		$\tau_{burst}$  (Gyr) & fixed  \\
		\hline
		\multicolumn{2}{c}{Delayed SFH}\\  
		\hline
		$age$ (Gyr) & 13	\\
		$\tau_{main}$  (Gyr) & 0.5, 1, 2, 3, 4, 5, 6, 7, 8, 9, 10, 20\\
		\hline
		\multicolumn{2}{c}{Trunc. delayed SFH}\\  
		\hline
		$age$ (Gyr) & \textbf{13}	\\
		$\tau_{main}$  (Gyr) & \textbf{0.5, 1, 2, 3, 4,}\\
					& \textbf{5, 6, 7, 8, 9, 10}, 12, 15, 20  	\\
		$age_{trunc}$ (Myr) & \textbf{0, 5, 10, 25, 50, 100, 150, 200, 250, 300,}\\
		 			 &  \textbf{ 350, 400, 450, 500, 550, 600, 650, 700,} \\
					 & \textbf{750, 800}\\
		$r_{SFR}$ 	& \textbf{0., 0.01, 0.02, 0.03, 0.04, 0.05, 0.1, 0.15, 0.2, 0.25,}	\\ 
		 			& \textbf{ 0.3, 0.35, 0.4, 0.45, 0.5, 0.55, 0.6, 0.65,}	 \\
		 			&  \textbf{0.7, 0.75, 0.8, 0.85, 0.9, 0.95, 1}, 1.05, 1.1\\
		\hline
		\multicolumn{2}{c}{Dust attenuation}\\  
		\hline
		$E(B-V)_*$	&  0.05, 0.075, 0.1, 0.125, 0.15, 0.175, \textbf{0.2},	\\
						&  0.225, 0.25, 0.275, 0.3, 0.325, 0.35, 0.375, 0.4	\\
		\hline		
		\multicolumn{2}{c}{Dust template: \cite{Dale14}}\\  
		\hline
		$\alpha$\tablefootmark{b}	& 	2., 2.375,  2.5, 2.75, \textbf{3.0}, 3.25, 3.5, 3.75, 4.  \\
		\hline
		\# of models & 136890\\
		\hline
		\label{fitparam}
		\end{tabular}
		\tablefootmark{a}{A exponentially decreasing SFH plus a constant star formation burst.}
		\tablefootmark{b}{Parameter linked to the $S_{60}/S_{100}$ IR ratio.}
	\end{table}

	We have now a set of mock SEDs for which we know the exact parameters, including the values of $age_{trunc}$ and $r_{SFR}$.
	Using the SED fitting function of CIGALE, we run the code on the mock catalogue in order to compare the output parameters to the ones we used to create the mock SEDs.
	The parameters used to perform the SED fitting are also presented in Table~\ref{fitparam}.
	
\begin{figure}%[!h] 
  		\includegraphics[width=\columnwidth]{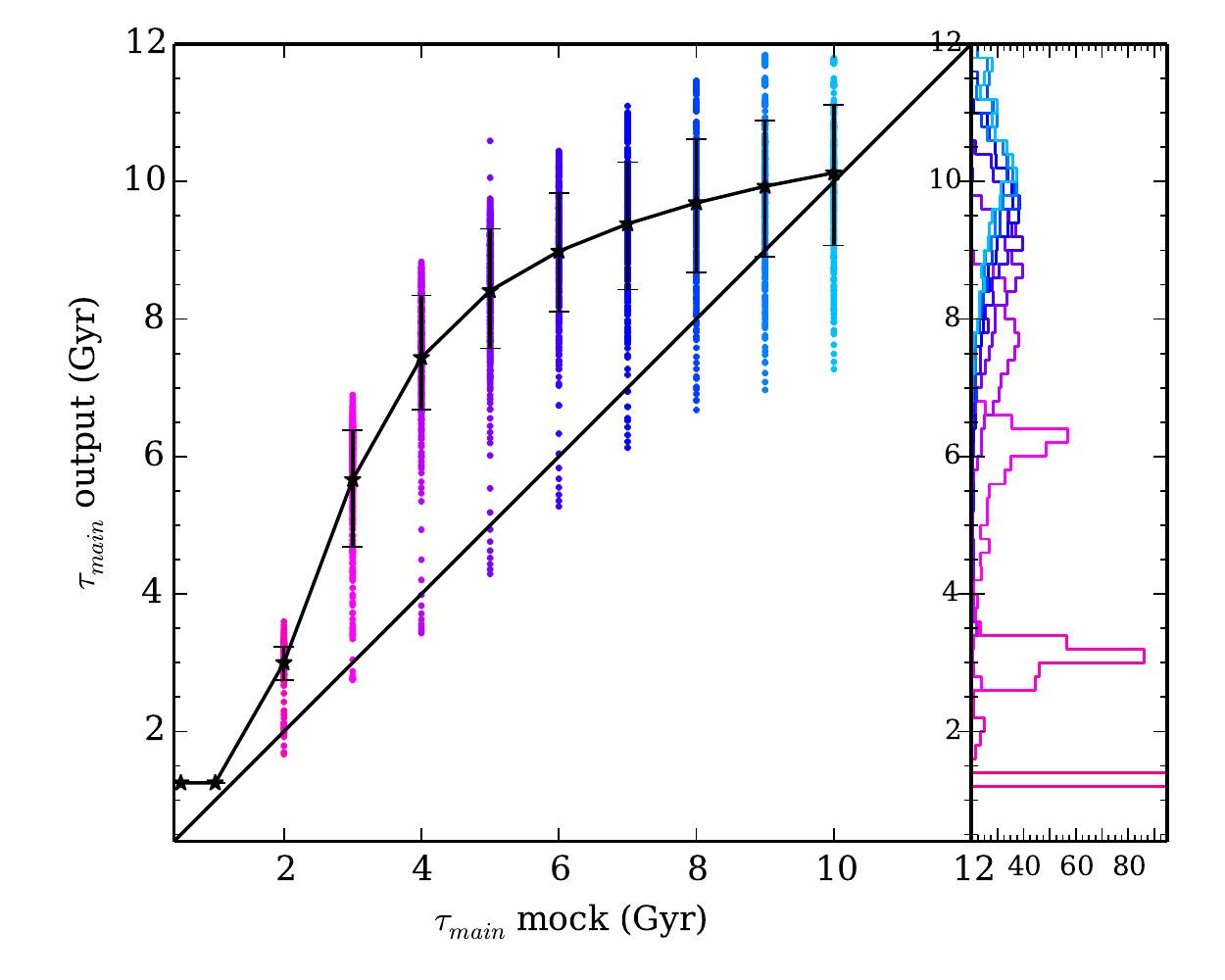}
  		\caption{ \label{mock1} Constraints on $\tau_{main}$ as a result of the mock catalogue analysis of the truncated SFH parameters. The dispersion of the output values are shown in blue, the mean and 64$^{th}$ percentile for each input values are marked in black. The black solid line is the one-to-one relationship. The right panel shows the distribution of the points for each input value of $\tau_{main}$.}
	\end{figure}

	Fig.~\ref{mock1} compares the output values of $\tau_{main}$ with the true values, used to create the mock galaxy catalogue.
	A perfectly recovered parameter would show a one-to-one relationship.
	Considering the known difficulty in recovering SFH parameters from SED fitting \citep[e.g.,][]{Giovannoli11}, the constraint on $\tau_{main}$ is relatively correct up to $\sim$4\,Gyr with an overestimate of a factor less than 2, but is degenerated for values higher than 4\,Gyr for which the code will provide an estimate between 8 and 10\,Gyr.
	Indeed, there is a flattening of the relation above 4\,Gyr.	
	For comparison, the mock analysis of $\tau_{main}$ used in the delayed SFH shows a slightly different behavior with values of $\tau_{main}<4$\,Gyr being very well constrained and higher values becoming overestimated up to 30\%. 
	It is however known that this parameter is difficult to constrain \citep[e.g.,][]{Buat14}.
	
\begin{figure}%[!h] 
  		\includegraphics[width=\columnwidth]{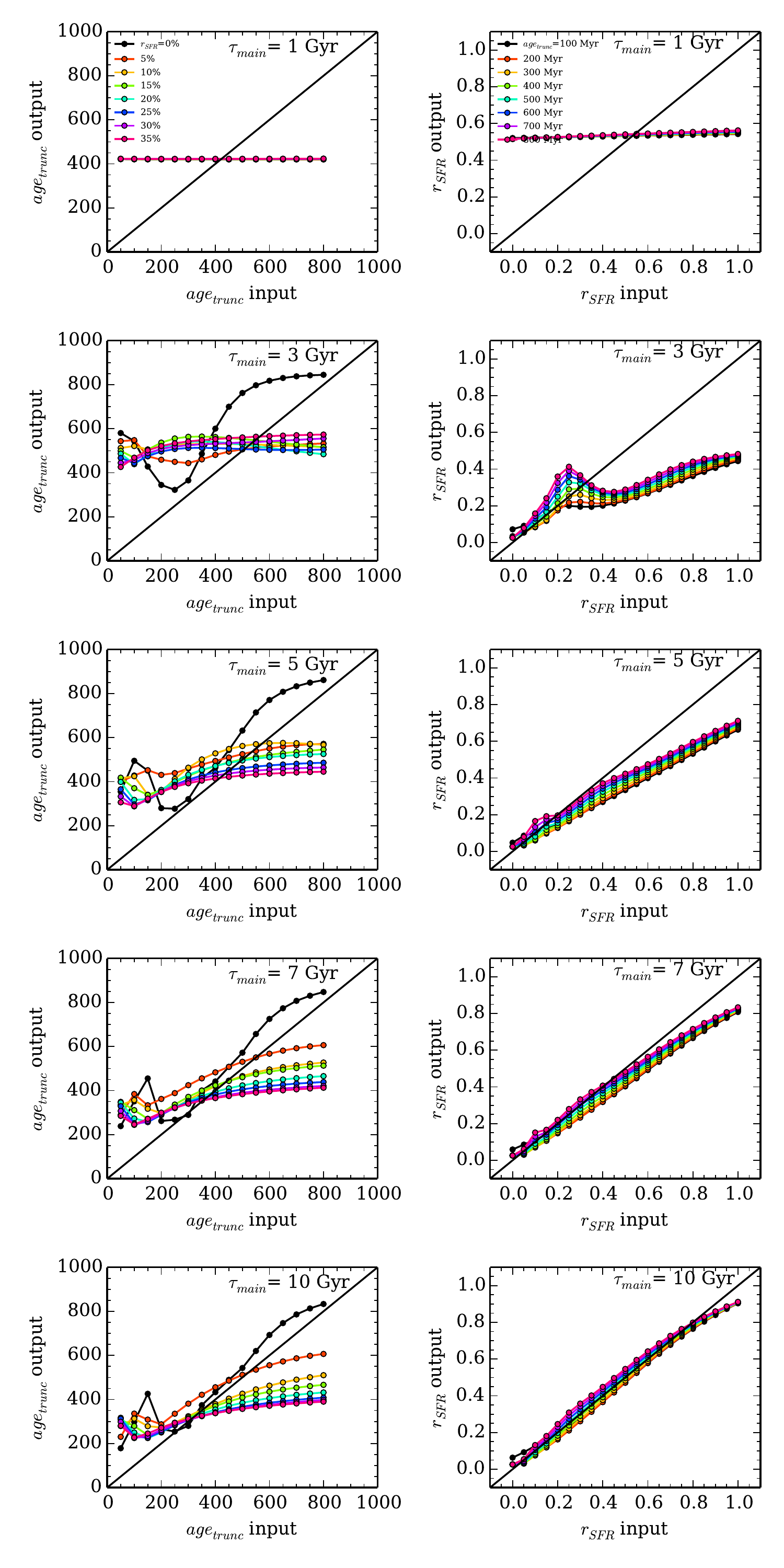}
  		\caption{ \label{mock2} Results of the mock catalogue analysis for the two parameters $age_{trunc}$ and $r_{SFR}$. The five left panels present the constraint on $age_{trunc}$ for five different $\tau_{main}$ (1, 3, 5, 7, and 10\,Gyr). The colored lines corresponds to different values of $r_{SFR}$. The five right panels display the constraint on $r_{SFR}$ for the same five values of $\tau_{main}$ (1, 3, 5,  7, and 10\,Gyr), the colored lines corresponding to different $age_{trunc}$. On each panel, the black solid line is the one-to-one relationship.}
	\end{figure}

	In a similar way to Fig.~\ref{mock1}, Fig.~\ref{mock2} presents the results for $age_{trunc}$ (top panels) and $r_{SFR}$ (bottom panels).
	The results are shown for different values of $\tau_{main}$ (1, 3, 5, 7, and 10\,Gyr).
	The  $age_{trunc}$ parameter is never constrained when $\tau_{main}=$1\,Gyr, and only for  $r_{SFR}=0$ for higher values of $\tau_{main}$.
	Indeed, as shown in Fig.~\ref{models}, the spectral range where $age_{trunc}$ impacts the SED is limited to 0.1-0.5\microns, where the $r_{SFR}$ also plays a role in the shape of the SED.
	Without any additional information, it is thus difficult for the model to determine precisely $age_{trunc}$ when $r_{SFR}>0$, as we can directly see in Fig.~\ref{mock2}.
	
	The $r_{SFR}$ parameter is relatively well estimated for true values of $\tau_{main}\geq5$\,Gyr.
	However, the recovered value of $\tau_{main}$  is biased toward higher values relative to the true value.
	As seen on Fig.~\ref{mock1}, a true value of $\tau_{main}\geq5$\,Gyr corresponds to a recovered value of $\sim$8\,Gyr.
	For low values, $\tau_{main}=1$\,Gyr, the results from the mock analysis show a flat relation meaning that $r_{SFR}$ is not constrained.
	Indeed, delayed SFH with a small $\tau_{main}$ of 1\,Gyr corresponds to an early rise of the SFR at early cosmic time followed by a rapid but smooth decrease of the SFR that is close to 0 at $t=13$\,Gyr.
	It is thus difficult to point out a rapid quenching of star formation in galaxies undergoing a smooth and long decreasing of their SFH over several Gyr.
	For higher $\tau_{main}$, the estimate is relatively close to the true value, especially for low values of $r_{SFR}$.
	For higher values of $r_{SFR}$, the higher $\tau_{main}$, the better the constraint.
	
	Thus, with the available set of filters, $\tau_{main}$ and $r_{SFR}$ can be relatively well constrained for $\tau_{main}\geq5$\,Gyr.
	To constrain $age_{trunc}$ from SED fitting, additional information is needed such as the $H\alpha$ flux density for instance.
	Indeed, the intensity of the $H\alpha$ lines is directly linked to the number of Lyman continuum photons and this could put a strong constraint on $age_{trunc}$ as seen from Fig.~\ref{models}, bottom panel \citep[e.g.,][]{Lee09,Weisz12,Boselli15}.
	This is the topic of a following study (Boselli et al., in prep).
	
%=================================================================================

\section{\label{sedfit}Application to the HRS galaxies}

	\begin{table}
		\centering
		\caption{Mean $\chi_{red}^2$ obtained using the three SFHs mentioned in this work and for three bins of \HIdef.}
		\begin{tabular}{l c c c c }
	 	\hline\hline
				 		 	& \#	galaxies&  2-exp-dec	& normal		& trunc.\\ 
		\hline
		\HIdef $<$ 0.3 			& 114	& 1.38	&	1.56	&	1.62	\\
		0.3 $\geq$ \HIdef $<$ 0.7 & 61		& 1.43	&	1.70	&	1.06 	\\
		\HIdef $\geq$ 0.7 		& 51		& 3.96	&	4.34	&	1.41	\\ 
		\hline		
		\label{chi2}
		\end{tabular}
	\end{table}

	\begin{figure}%[!h] 
  		\includegraphics[width=\columnwidth]{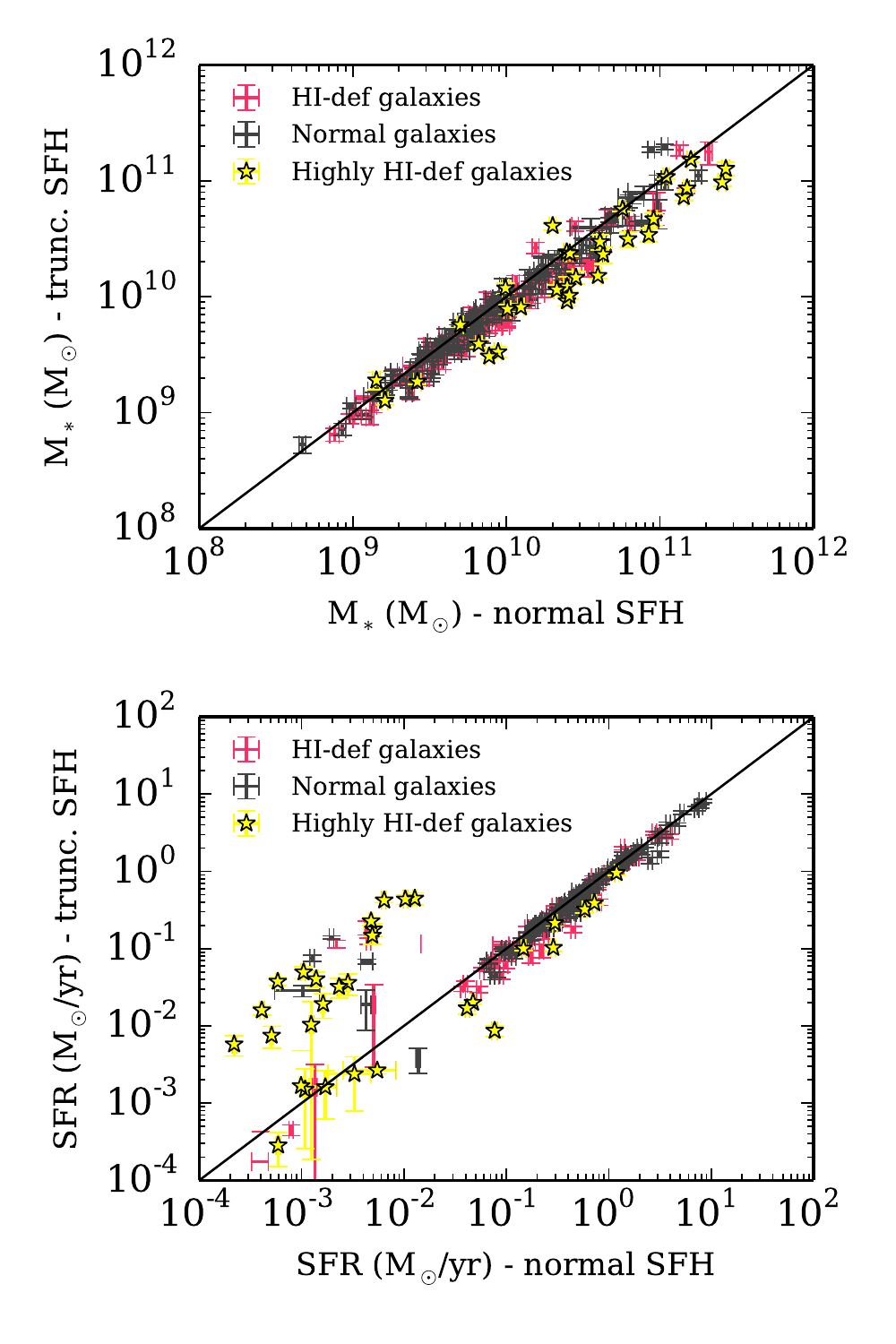}
  		\caption{ \label{comp_mstar_sfr} Comparison between the stellar masses (top panel) and SFR (bottom panel) obtained using the truncated and normal delayed SFH. Normal galaxies are shown in grey whereas \HI-deficient, and highly deficient galaxies (with \HIdef$\geq$1) are shown in red and yellow, respectively. The block solid lines represent the one-to-one relationship.}
	\end{figure}

	%=========
	\subsection{Results from SED fitting}
	
	We run CIGALE on the subsample of HRS late-type galaxies, including 136 normal and 92 \HI-deficient galaxies, using the truncated delayed SFH for all of them.
	The fits are performed over the entire spectrum from the UV to the submm.
	Indeed, benefiting from the philosophy of CIGALE which is based on energy balance between UV-optical and IR, IR and submm data provide an additional constraint on the dust attenuation.
	The set of parameters used for the fit are presented in Table~\ref{fitparam}.
	
	The mock analysis showed that the $age_{trunc}$ parameter is not constrained from broad band photometry.
	Nevertheless, we perform a first run of CIGALE leaving this parameter free.
	The resulting distribution of  $age_{trunc}$ is indeed peaking at the same value for both the normal and deficient subsamples and with a similar spread, confirming the difficulty to constrain this parameter (see Fig.~\ref{hist_agetruncfree}).
	We thus decide to fix its value to $350$\,Myr and perform a second run with this parameter fixed.
	From now on, we present and discuss the results of the run with fixed $age_{trunc}$.
	
	For a normal galaxy, the quality of the fit in UV-optical is the same compared to a delayed SFH and a 2-exponentially decreasing SFH (Fig.~\ref{sedex}, top panel).
	For the \HI-deficient galaxies, the truncated SFH results in a better agreement between the modeled SEDs and the data, both the FUV and NUV flux densities are reproduced by the model (Fig.~\ref{sedex}) where other SFH assumptions failed to do so.
	In addition, the computed models are able to reproduce the emission of the entire stellar populations, both young and old stars.
	Although it is complicated to compare different $\chi_{red}^2$ obtained from models with different degrees of freedom, we give in Table~\ref{chi2} the mean $\chi_{red}^2$ obtained for the three SFH mentioned in this work and for three bins of \HIdef. 
	The mean $\chi_{red}^2$ obtained with normal SFHs is $\sim$3 times higher for the \HIdef\ galaxies whereas the truncated SFH provides consistent $\chi_{red}^2$ for all of the three subsamples.
	
	The use of the truncated SFH compared to the normal delayed SFH mostly impacts the UV and NIR providing better fits (Fig.~\ref{sedex}). 
	These domains being crucial to determine the SFR and stellar mass of galaxies, we show in Fig.~\ref{comp_mstar_sfr} the differences obtained on these parameters using the truncated and normal delayed SFH.
	For the normal galaxies, the comparison shows a small scatter but this is expected as different SFHs are used \citep[e.g.,][]{Pforr12,Buat14,Ciesla15}.
	This small scatter increases when we consider the \HI-deficient galaxies leading to lower stellar masses obtained with the truncated SFH and smaller SFRs.
	The effect is more striking when considering the most deficient galaxies (with \HIdef$\geq$1) where the stellar mass is overestimated by 30\%  and the SFR by a factor of $\sim$15 on average when using a normal delayed SFH instead of the truncated SFH.
	The large difference observed in the SFR estimates can be understood from Fig.~\ref{sedex} where it is shown that the normal delayed SFH strongly underestimates the UV emission of the \HI-deficient galaxies.
	This emphasizes the importance of using an appropriate SFH to retrieve the physical parameters of galaxies.

\begin{figure}%[!h] 
  		\includegraphics[width=\columnwidth]{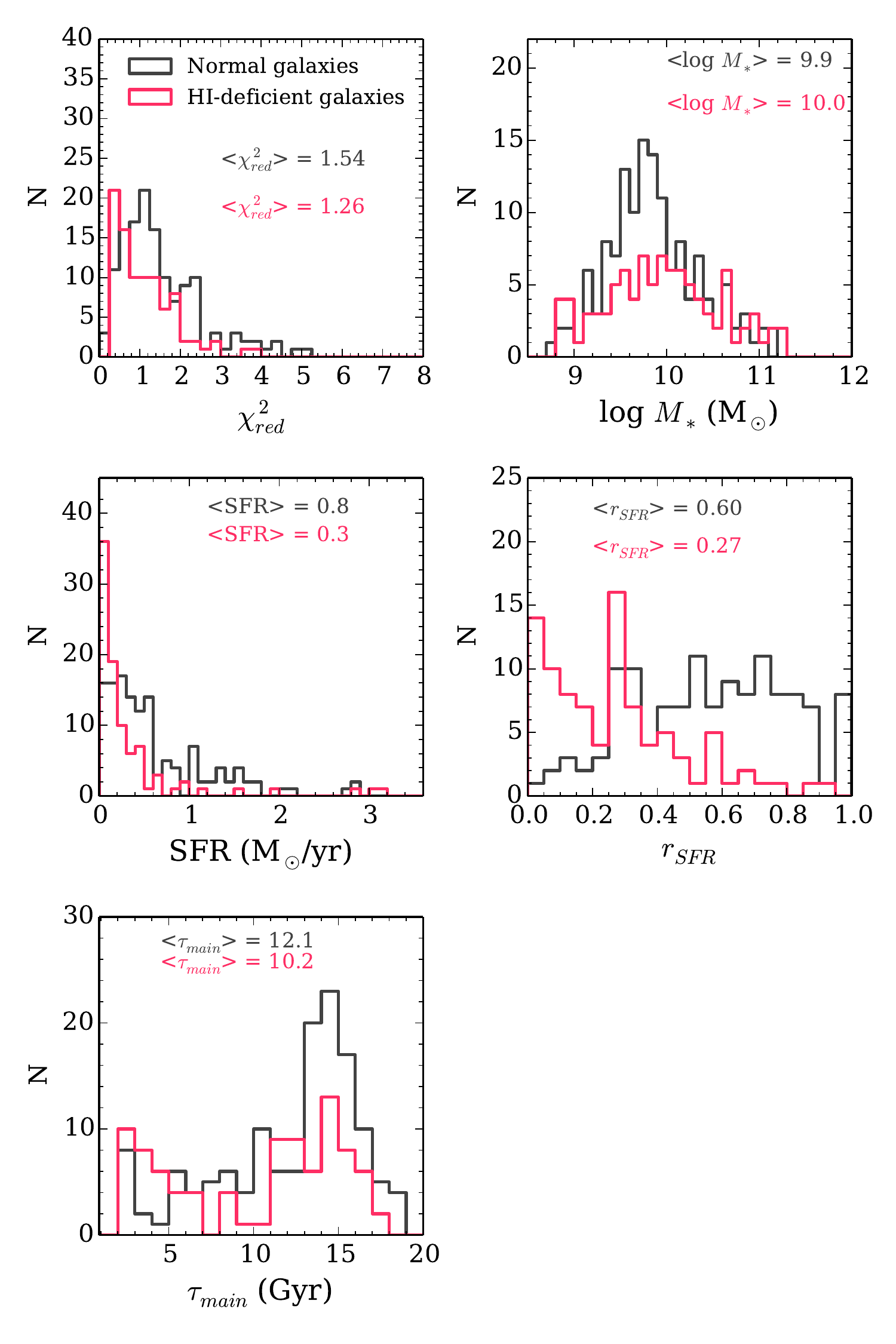}
  		\caption{ \label{hist} Distribution of the output parameters obtained from the SED fitting procedure with CIGALE. The results for the normal galaxy sample are shown in grey while the results for the \HI-deficient sample are shown in red. }
	\end{figure}

	The distribution of the $\chi^2$, $M_*$, SFR, $r_{SFR}$, and $\tau_{main}$  obtained from the SED fitting using the truncated SFH are presented in Fig.~\ref{hist} for both normal and \HI-deficient galaxies.
	The $\chi^2$ distributions of the normal and \HI-deficient samples peak at different values, with a mean of 1.54 and 1.26, respectively.
	The distributions of the stellar mass of both subsamples are similar, peaking at $\log M_* \approx9.9-10.0$ and spreading from 8.5 to 11.2.
	However, the distribution of star formation rates of the \HI-deficient clearly shows that most of these galaxies have a very low, almost zero, SFR as expected for quenched galaxies.
	The distribution of the $r_{SFR}$ parameter shows two different behaviors for normal and deficient galaxies.
	The \HI-deficient subsample distribution shows lower values with 34 sources with $r_{SFR}<0.15$ out of 92 deficients galaxies.
	There is a second peak at 0.3 in the $r_{SFR}$ distribution with 16 galaxies.
	The $r_{SFR}$ distribution of normal galaxies is flat and shifted toward larger values although not as close to 1 as one would expect.
	This can be explained from Fig.~\ref{mock2} where we see that, even for large values of $\tau_{main}$, $r_{SFR}$ values close to 1 tend to be underestimated by at least 10-20\%.
	Finally the distribution of $\tau_{main}$ shows that most of the sources have $\tau_{main}\geq5$\,Gyr, the typical value above which the uncertainty on $r_{SFR}$ begins to reduce, as discussed in Sect.~\ref{cons}.  
	
%=================================================================================
	\subsection{\label{relation}Relation between the strength of the quenching and the \HI-deficiency}	
	
\begin{figure}%[!h] 
  		\includegraphics[width=\columnwidth]{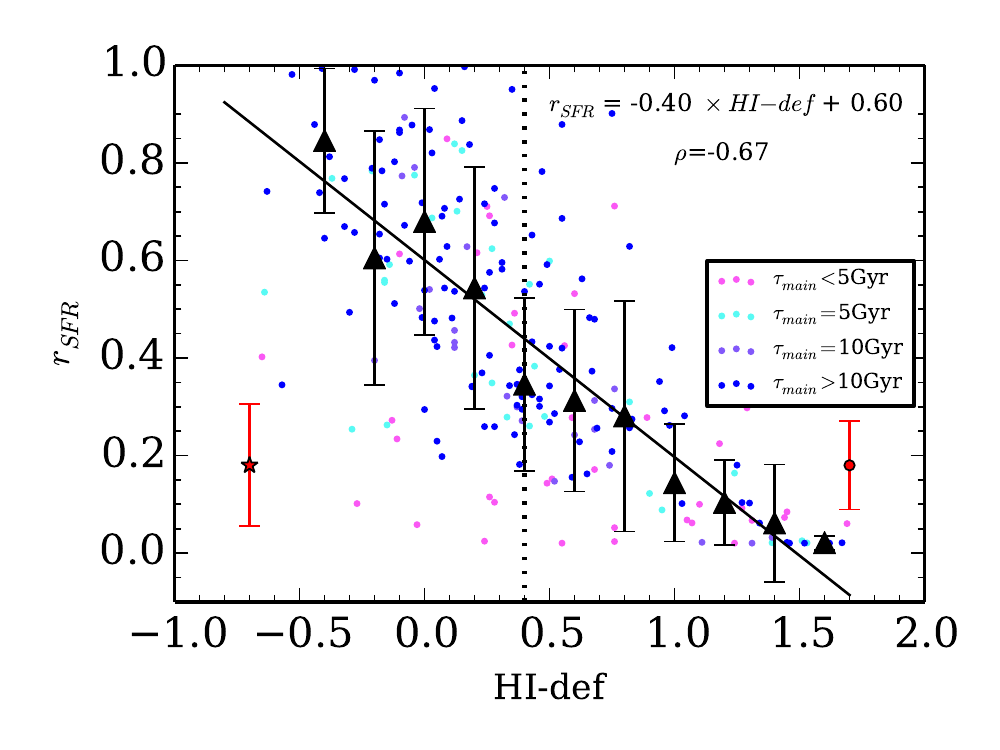}
  		\caption{ \label{rsfrvsdefhi} Relation between $r_{SFR}$ and the \HI-deficiency. Data points are color coded according to the value of $\tau_{main}$ obtained from the fit. The mean errors bars for each subsample are provided with the red star marker for the normal galaxy sample and with the red filled circle for the deficient galaxies. The Spearman correlation coefficient $\rho$ of the relation is indicated.  Black triangles are the median values in bins of $\Delta$\,\HIdef $= 0.2$, the error bars are the standard deviation of the points in each bins. The black filled line is best linear fit to the data. The black dotted line indicates the adopted threshold used to separate normal from deficient galaxies.}
	\end{figure}

	To further examine the $r_{SFR}$ parameter, we show in Fig.~\ref{rsfrvsdefhi} the relation between this parameter and \HIdef.	
	With a Spearman correlation coefficient of $-0.67$, there is a good anti-correlation between these two parameters for which the best linear fit results in:
	
	\begin{equation}
	r_{SFR} = -0.40 \times HI-def + 0.60.
	\end{equation}
	
	Indeed, galaxies with a high \HIdef\ parameter are the ones most affected by the cluster environment with a large fraction of their gas content stripped. 
	These sources are thus the most quenched and, consequently, have a very low $r_{SFR}$ value.
	On the contrary, sources with negative values of  \HIdef, with no gas stripped, show high values of $r_{SFR}$.
	However, we would have expected values closer to 1 for normal galaxies and less dispersion.
	A value lower than 1 for the normal star forming galaxies means that the assumption of a quasi constant SFR over the last few hundreds of Myr from the delayed SFH is too strong.
	Thus high values of $r_{SFR}$, although not equal to 1, may model a small decrease towards the end of the evolution additional to the one obtained with the simple smooth delayed SFH.
	From Fig.~\ref{rsfrvsdefhi}, we see that normal galaxies having the lowest $r_{SFR}$ also have a low value of $\tau_{main}$.
	From the results of the mock catalogue analysis, low values of $\tau_{main}$ lead to an underestimate of  $r_{SFR}$.
	To understand if the values lower than 1 obtained for the normal galaxies are due to the underestimate of high values of $r_{SFR}$, as seen in the mock catalogue results (Fig.~\ref{mock2}), we derived $r_{SFR}$ corrections from the mock results and applied them to the $r_{SFR}$ values.
	The relation observed on Fig.~\ref{rsfrvsdefhi} remains globally unchanged, with a Spearman correlation coefficient of $-0.66$.
	This is due to the fact that most of the sources have a $\tau_{main}$ value larger than 5\,Gyr and thus small corrections.

	This relation between the two parameters can be useful as the \HIdef\ parameter, quantifying the impact of the environment on the gas content of a galaxy, and even gas measurements are not available for a large number of sources, especially at high redshifts.
	The NUV-r color, excellent proxy for gas content in galaxies, also shows a good correlation with \HIdef\ but slightly more dispersed, with a Spearman correlation coefficient of $0.61$ (Fig.~\ref{nuvrdef}).
	From SED fitting, with a large photometric coverage, especially including the rest frame UV, it would be possible to obtain an estimate of this parameter.
	More generally, it means that broad band SED fitting can provide information on SFH that are recently perturbed, such as a rapid decrease of the star formation activity.

%=================================================================================
	\subsection{\label{msd}Position of the \HI-deficient galaxies on the SFR--$M_*$ diagram}
	
\begin{figure}%[!h] 
  		\includegraphics[width=\columnwidth]{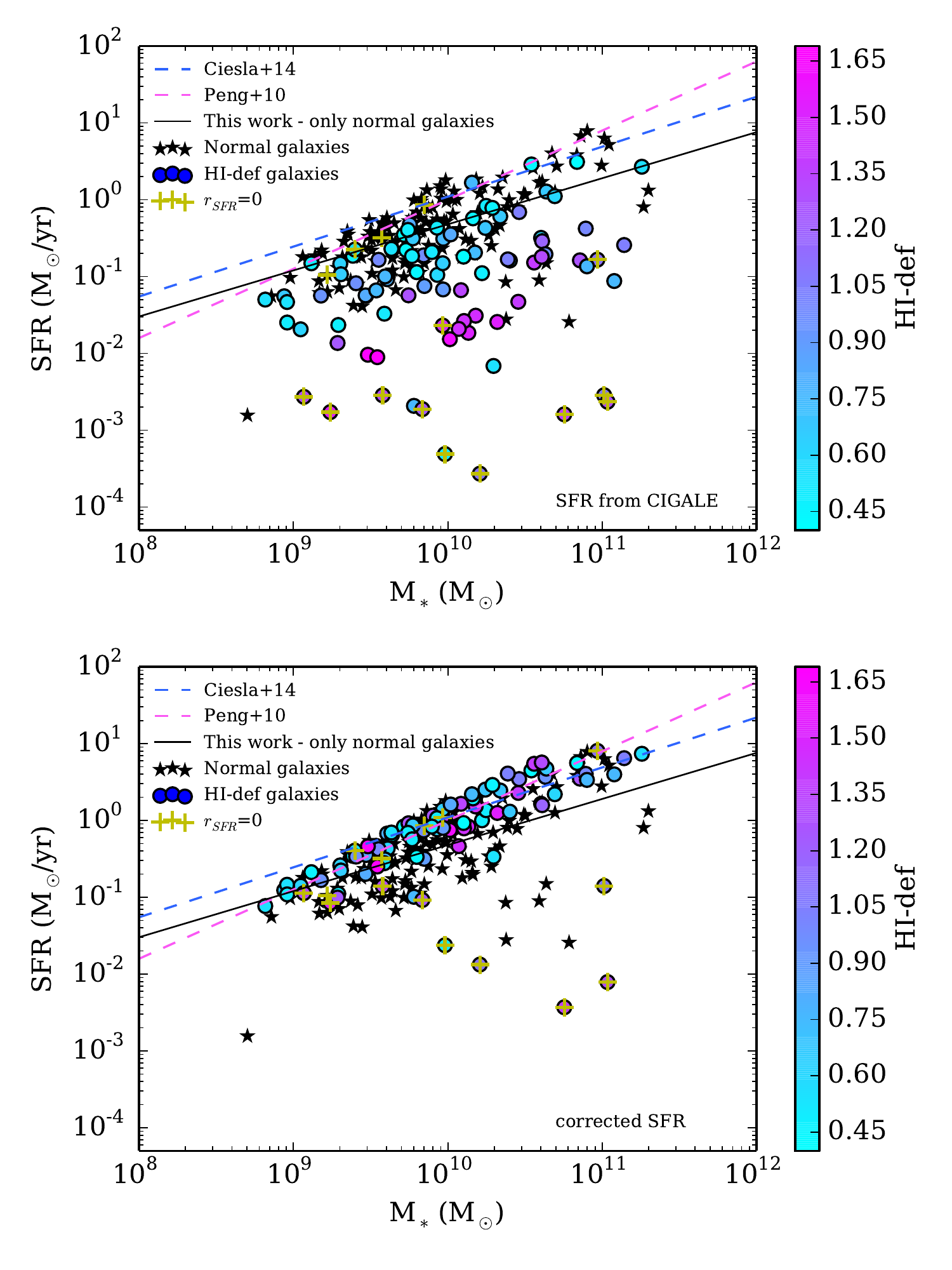}
  		\caption{\label{ms}Relation between the SFR and $M_*$ of the HRS subsamples. Black stars are the normal galaxies, filled circles are the \HI-deficient galaxies color coded according to the \HIdef\ parameter. Yellow crosses point out the sources for which $r_{SFR}$=0. Main sequence relations found at $z=0$ are shown in dashed lines: in blue the one from \cite{Ciesla14}, in red the relation of \cite{Peng10}, the relation derived in this work is the black solid line. Top panel: observed relation obtained with CIGALE. Bottom panel: estimated positions of the \HI-deficient galaxies before interaction with the environment obtained after correcting their instantaneous SFR with the $r_{SFR}$ parameter. }
	\end{figure}

	Recent studies of large samples of galaxies showed that the majority of star-forming galaxies follow a SFR--$M_*$ correlation, called the main sequence \citep[MS,][]{Noeske07_SFseq,Elbaz07,Peng10,Speagle14}.
	Because the sample was selected in K-band to be complete in mass, the HRS galaxies are ideal to probe the MS at $z=0$.
	In Fig.~\ref{ms}, we show the positions of our normal star-forming and \HI-deficient subsamples on the MS diagram.
	In this work, stellar masses and SFR are estimated through our SED fitting procedure.
	The normal galaxies are the same as those studied in \cite{Ciesla14} who demonstrated that, using different estimates of SFRs and $M_*$, the gas rich late-type subsample of HRS galaxies lie on the MS derived by \cite{Peng10}, as shown in Fig.~\ref{ms}, but with a slightly flatter slope.
	The new estimates of the masses and SFRs now place the galaxies slightly lower than the MS derived in \cite{Ciesla14} due to the different methods employed to determine them.
	At lower masses, the HRS galaxies agree better with the relation determined by \cite{Peng10}, even though the slope of the MS is still flatter.
	With SFRs determined from new $H\alpha$ imaging, \cite{Boselli15} showed that the best linear fit to their SFR-$M_*$ relation is closer to \cite{Ciesla14} whereas the bisector of the relation is closer to the relation of \cite{Peng10}.
	As the slope of the MS is sensitive to the methods and assumptions made to derive the SFRs and stellar masses, we do not discuss further the different relations obtained by \cite{Peng10}, \cite{Ciesla14}, and \cite{Boselli15} because the results of this work are not linked to the MS slope.
	
	In Fig.~\ref{ms} (top panel), the galaxies with an \HIdef\ lower than $\sim$1.00 seem to lie at the lowest part of the MS compared to the normal galaxies, and the most extreme deficient ones (\HIdef$>$1.00) are off the MS with very low SFR compared to their stellar mass.
 	Indeed, previous studies showed that the \HI-deficient galaxies lie in the green valley and are considered as intermediates between star-forming disks and passive early-type objects \citep[e.g.,][]{Boselli08}.
	Furthermore, \cite{Boselli15} determined the SFR-$M_*$ relation for the normal and the \HI-deficient star forming galaxies and found a shift of 0.65\,dex between the two relations.
	However, using the observed SFR of these sources and the estimate of the $r_{SFR}$ parameter obtained from the SED fitting, we can have an estimate of the SFR of the galaxies before being affected by the dense environment of the Virgo cluster.
	In Fig.~\ref{ms} bottom panel, we show the positions of the \HI-deficient sources using their ``corrected'' SFR.
	After correction, almost all of the \HI-deficient galaxies lie on the MS formed by the normal star-forming galaxies of the HRS sample.
	Five deficient sources remain below the MS.
	These five sources have a $r_{SFR}$ equal to 0 and a very low SFR with $SFR/error$ lower than 1.6.
	Given the definition of this parameter, it is not possible to obtain an estimate of the SFR of these sources before the truncation.
	The properties of the \HI-deficient galaxies before being affected by the cluster environment are thus consistent with those of isolated unperturbed star forming galaxies.
	
%=================================================================================
\section{\label{discussion}Discussion}	

\begin{figure}%[!h] 
  	\includegraphics[width=\columnwidth]{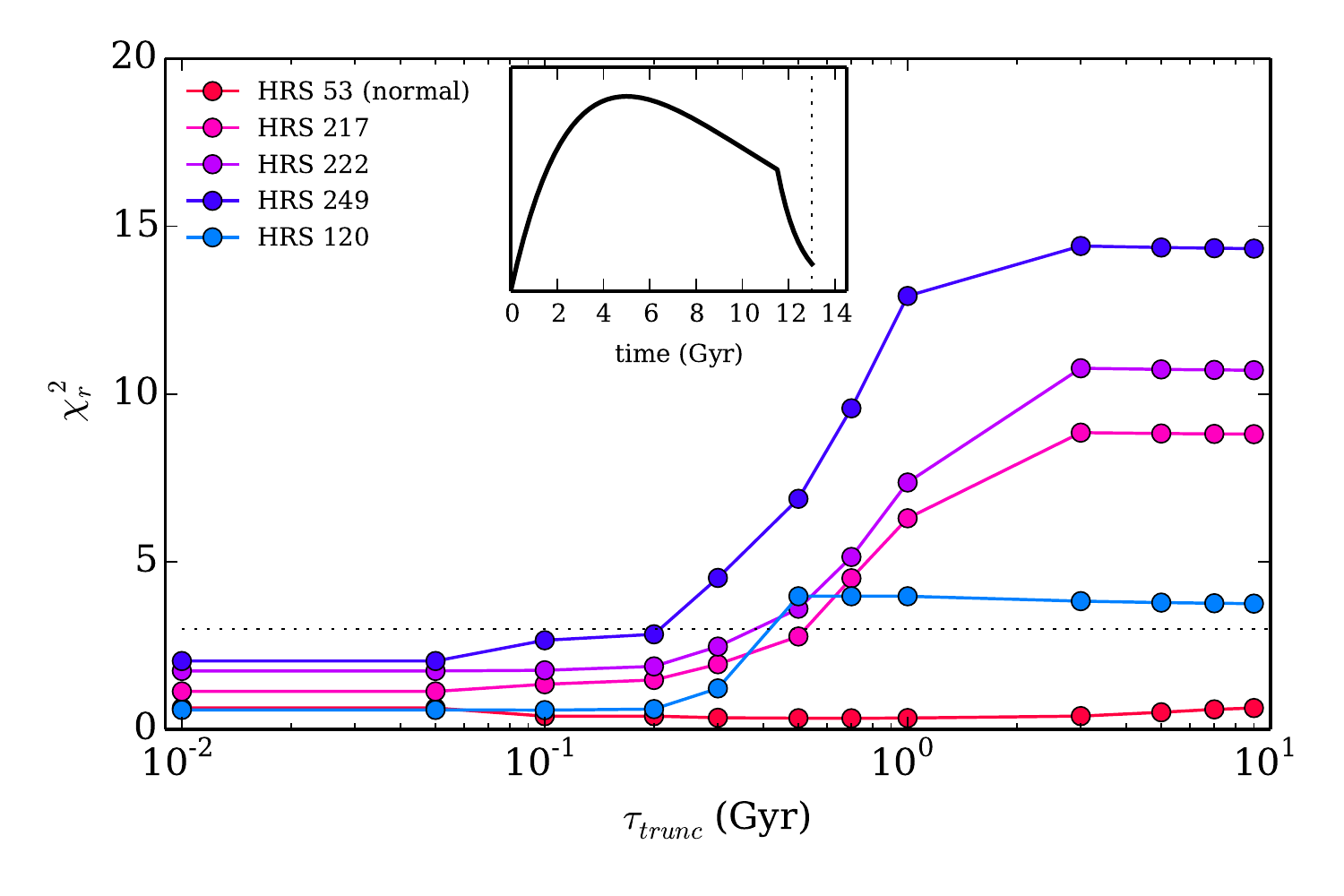}
  	\caption{\label{tau} Variations of the $\chi^2$ as a function of $\tau_{trunc}$ for the five galaxies presented in Fig.~\ref{sedex}. The horizontal dotted line shows a $\chi^2$ of 3 below which fits are usually considered as good. The insert panel shows the shape of the SFH considered in this test.}
\end{figure}

In this work, we modeled the SFH of galaxies undergoing a rapid quenching by allowing the possibility of a sudden truncation in the SFR.
To limit the possible degeneracies that could occur between the parameters used to model the galaxy emission, we restricted ourselves to an instantaneous break in the SFH.
This assumption is strong as even if ram pressure affects the gaseous component of the galaxies in a few Myr, the time to completely remove the gas and stop the star formation activity depends on several parameters such as the mass of the galaxy and its dynamics in the cluster, and is estimated by previous studies to be $<$1.5\,Gyr.
To test if the assumption of a smooth decrease of the SFR rather than an  instantaneous one affects our results, we implemented in the SFH the possibility of an exponential decrease of the SFR after $t_{trunc}$ for which the e-folding time, $\tau_{trunc}$, is a free parameter.
In Fig.~\ref{tau}, we present the $\chi^2$ values obtained varying $\tau_{trunc}$ for the five galaxies presented in Fig.~\ref{sedex}.
A high value of $\tau_{trunc}$ corresponds to a normal delayed SFH without any truncation and small values of $\tau_{trunc}$ to a rapid decrease of the SFH.
For the four deficient sources, there is a strong decrease of the $\chi^2$ values at a specific range of $\tau_{trunc}$ between $\sim$200 and 500\,Myr.
In other words, the $\chi^2$ is almost constant and below 3, for $\tau_{trunc}$ lower than 200-500\,Myr, depending on the galaxy.
In this 200-500\,Myr range, the $\chi^2$ drops from values between 4 and 15 to be less than 3 for lower values of $\tau_{trunc}$.
The results of this test implies that drastic star formation quenching occurring in $\leq$500\,Myr can be modeled by an instantaneous drop of the star formation activity.

In Sect.~\ref{msd}, we correct the SFR of the \HI-deficient galaxies to show that they were once on the main sequence relation.
However, even unperturbed objects have $r_{SFR}$ that is not exactly equal to 1 but scattered in the 0.7-1 range.
These high values of  $r_{SFR}$ translate into a slow decrease of their SFR.
We did not correct the SFR of these sources as we know from their \HI-deficiency that they did not undergo a rapid decrease of their star formation activity.
As we discuss in Sect.~\ref{relation}, the \HIdef\ parameter is not available for a lot of sources and catalogues.
Thus some other criteria and associated methodology should be applied in order to identify systems which underwent a rapid decrease of their star formation activity and estimate their SFR prior to quenching.
As shown on Fig.~\ref{rsfrvsdefhi}, the mean value of  $r_{SFR}$ at \HIdef=0.4, the typical value separating the gas-rich from the \HI-deficient galaxies, is 0.3.
Thus one possibility is to consider as quenched sources with $r_{SFR}\leq0.3$.
Another possibility is to decide an arbitrary cut in the ratio between the $\chi^2$ obtained for the delayed and truncated SFH to estimate which of the two SFH better models the SED.

\begin{figure}%[!h] 
  	\includegraphics[width=\columnwidth]{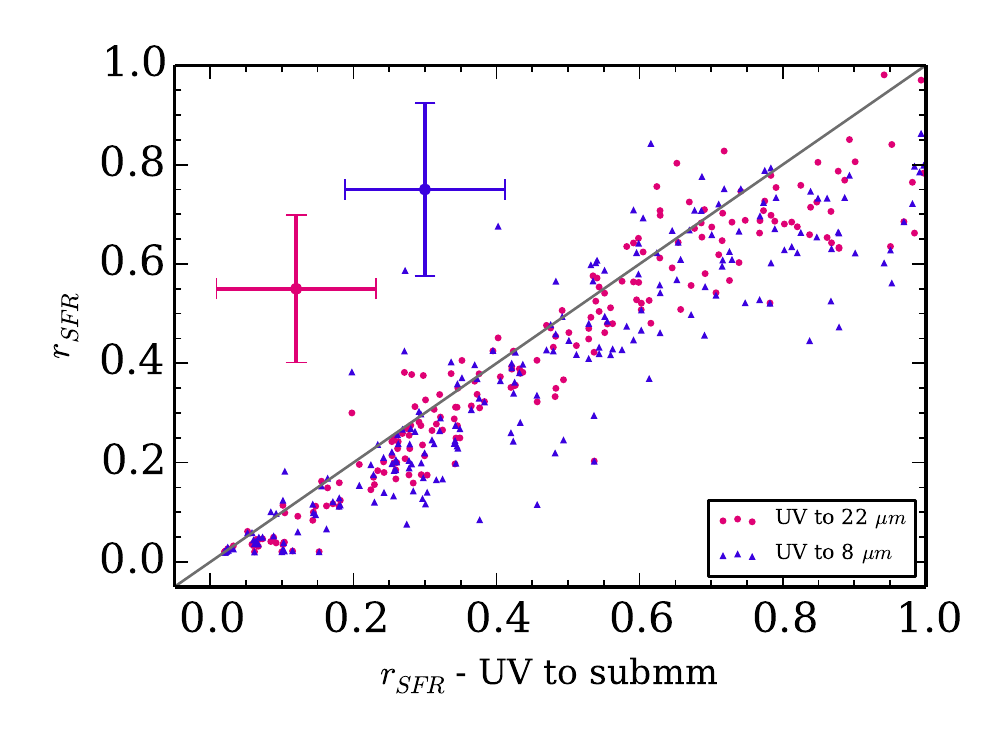}
  	\caption{\label{noIR} Impact of the photometric coverage on the derivation of the $r_{SFR}$ parameter. Pink points shows the  relation between the the values obtained when the photometric coverage stops at 22\microns\ and the values obtained from a full coverage. The purple points show the relation when the photometric coverage stops at 8\microns. Mean error bars are shown for both subsets.  }
\end{figure}

The HRS galaxies benefit from a wealth of photometric data from UV-to-submm. 
However, at higher redshifts, IR observations are not always available and the photometric coverage of these sources often stops with WISE or \textit{Spitzer}/MIPS 22-24\microns\ data or even at \textit{Spitzer}/IRAC wavelengths. 
We show in Fig.~\ref{noIR} the relation between the $r_{SFR}$ values obtained with a full photometric coverage and the values obtained stopping at 22\microns\ and then at 8\microns\ rest frame.
There is a weak underestimate of the parameter in absence of IR data as well as an increase of the dispersion.
However the differences are still smaller than the error bars.
We notice that the estimates of low $r_{SFR}$ ($<0.2$) values are consistent with the one obtained with the full coverage.
The dispersion is however higher for larger $r_{SFR}$ values.
We conclude from this test that the results of this work hold for source for which the IR domain is not well covered. 
	
\begin{figure}%[!h] 
  	\includegraphics[width=\columnwidth]{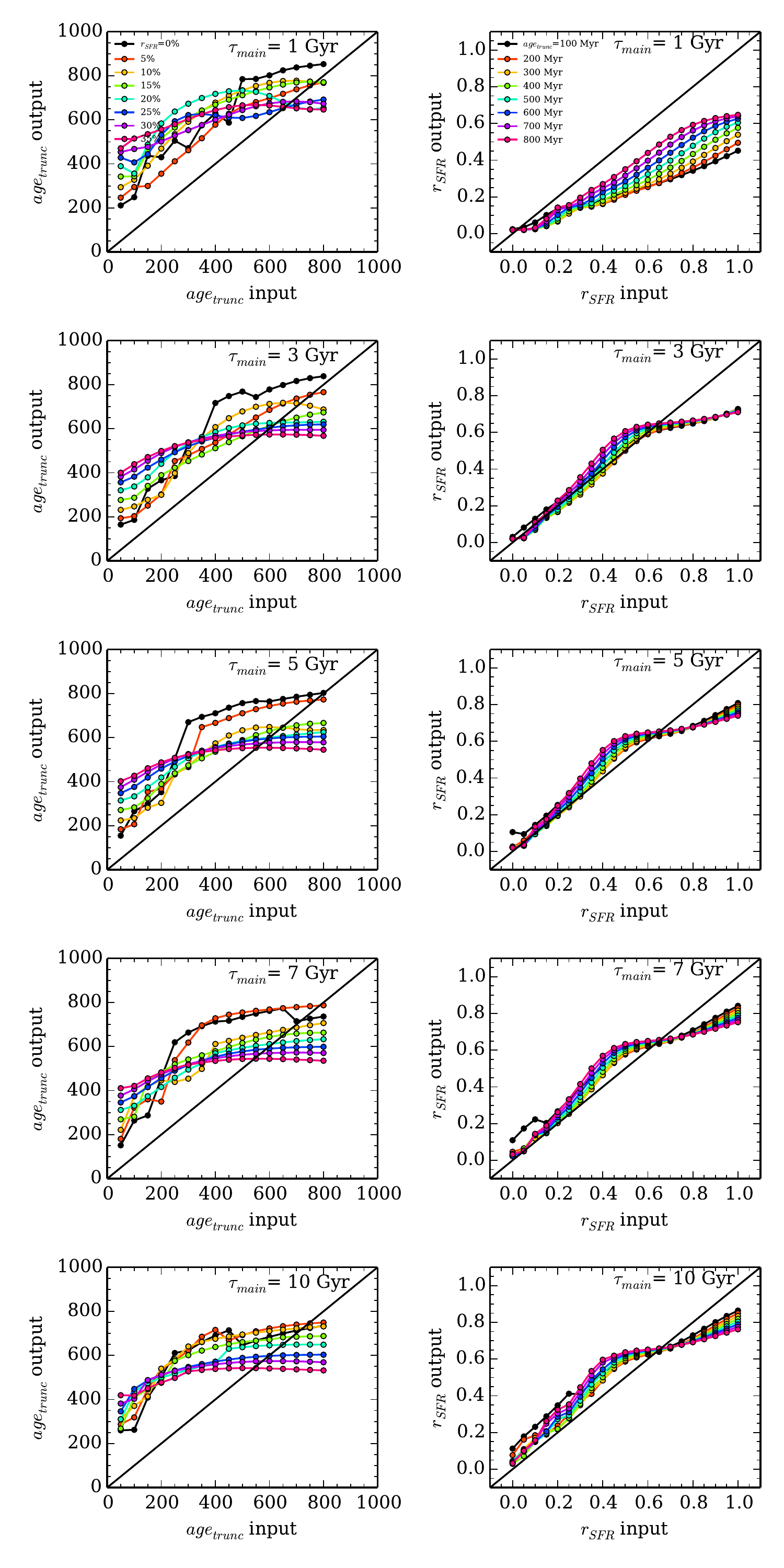}
  	\caption{\label{mockhighz} Results of the $z=2$ mock catalogue analysis for the two parameters $age_{trunc}$ and $r_{SFR}$. The five left panels present the constraint on $age_{trunc}$ for five different $\tau_{main}$ (1, 3, 5, 7, and 10\,Gyr). The colored lines corresponds to different values of $r_{SFR}$. The five right panels display the constraint on $r_{SFR}$ for the same five values of $\tau_{main}$ (1, 3, 5,  7, and 10\,Gyr), the colored lines corresponding to different $age_{trunc}$. On each panel, the black solid line is the one-to-one relationship.}
\end{figure}

We test our method on local galaxies with a relatively low dust attenuation.
However, in the redshift range $z=1--3$, UV galaxy emission is deeply obscured by dust absorption.
To evaluate the ability of our SED fitting method to identify star formation quenching at these redshifts, we followed the method presented in Sect.~\ref{cons} to build a mock catalogue of $z=2$ galaxies.
We assume $A_{FUV}=4$\,mag as a mean attenuation at redshift 2, determined by \cite{Buat15}.
The results of the mock analysis are presented in Fig.~\ref{mockhighz}.
As for $z=0$ galaxies, the $age_{trunc}$ parameter is not constrained, whatever the value of $\tau_{main}$.
However, with small variations around the one-to-one relationship, the $r_{SFR}$ parameter is relatively well recovered.
The main difference compared to the mock analysis of $z=0$ sources is the good constraint on $r_{SFR}$ when $\tau_{main}=3$\,Gyr.
The results of this test imply that the method proposed in this work can be applied to high redshift galaxies.

In this study, we discuss ram pressure stripping as an example of rapid quenching because we have a sample of local galaxies with sufficient photometric coverage and ancillary data to test the effect of this mechanism on the SED of galaxies.
However, even though we focused on one particular mechanism, the results of this work suggest than broad band SED fitting is a powerful tool to identify peculiar SFH of galaxies such as rapid drop of the star formation activity. 
Indeed, SFH modeling a recent burst in the star formation activity are widely used in the literature \citep[e.g.,][]{Papovich01,Borch06,Gawiser07,Lee09,Buat14}, however we show in this work that a rapid decrease can also be identified.

%=================================================================================
\section{Conclusions}

We defined a truncated delayed SFH to model the SEDs of galaxies that underwent a rapid quenching of their star formation activity.
Using the CIGALE SED fitting code, we showed that the ratio between the instantaneous SFR and the SFR just before the truncation of the SFH is well constrained as long as UV rest frame data are available.

This SED fitting procedure is applied to the \textit{Herschel} Reference Survey (HRS) as it contains both isolated galaxies and sources lying in the dense environment of the Virgo cluster.
These objects are \HI-deficient due to ram pressure happening in the cluster.
We showed that the truncated delayed SFH manages to reproduce their UV-to-NIR SED while the usual SFH assumptions fail.
An anti-correlation is found between $r_{SFR}$ and \HIdef\, the parameter quantifying the gas deficiency of the Virgo galaxies,  with a Spearman correlation coefficient of $-0.67$, implying that SED fitting can be used to provide an tentative estimate of the gas deficiency of galaxies for which \HI\ observations are not available.
The HRS galaxies are placed on the SFR-$M_*$ diagram showing that the \HI-deficient sources lie in the quiescent region in agreement with what was found in previous studies.
Using the $r_{SFR}$ parameter, we derive the SFR of these sources before quenching and show that they were initially on the galaxy main sequence relation. 

We discussed the assumption made on an instantaneous break in the SFH and showed that it holds for quenching mechanisms affecting the star formation activity in less than 200-500\,Myr.
The estimate of  $r_{SFR}$ in absence of IR data is consistent with what obtained with a full photometric coverage within the error bars.
Furthermore, the truncated SFH proposed in this work can also be used for deeply obscured ($A_{FUV}\approx4$\,mag) high redshift sources. 
SED fitting is thus a powerful tool to identify galaxies that underwent a rapid star formation quenching and can provide a tentative estimate of their gas deficiency.

%=================================================================================
\begin{acknowledgements}
We thank the referee for his/her comments that helped improving the paper.
L.\,C. warmly thanks M.~Boquien, Y.~Roehlly and D.~Burgarella for developing the new version of CIGALE on which the present work relies on and L.~Cortese for useful comments.
L.\,C. benefited  from the {\sc  thales} project 383549 that  is jointly funded by the European Union and the Greek Government in the framework of the program ``Education and lifelong learning''.
The research leading to these results has received funding from the European Union Seventh Framework Programme (FP7/2007-2013) under grant agreement n° 312725.
\end{acknowledgements}

%%=================================================================================
\bibliographystyle{aa}
\bibliography{pcigale_def}

%=================================================================================
\appendix
\section{\label{agetruncfree}Results from SED fitting: $age_{trunc}$ as a free parameter}

\begin{figure}%[!h] 
  	\includegraphics[width=\columnwidth]{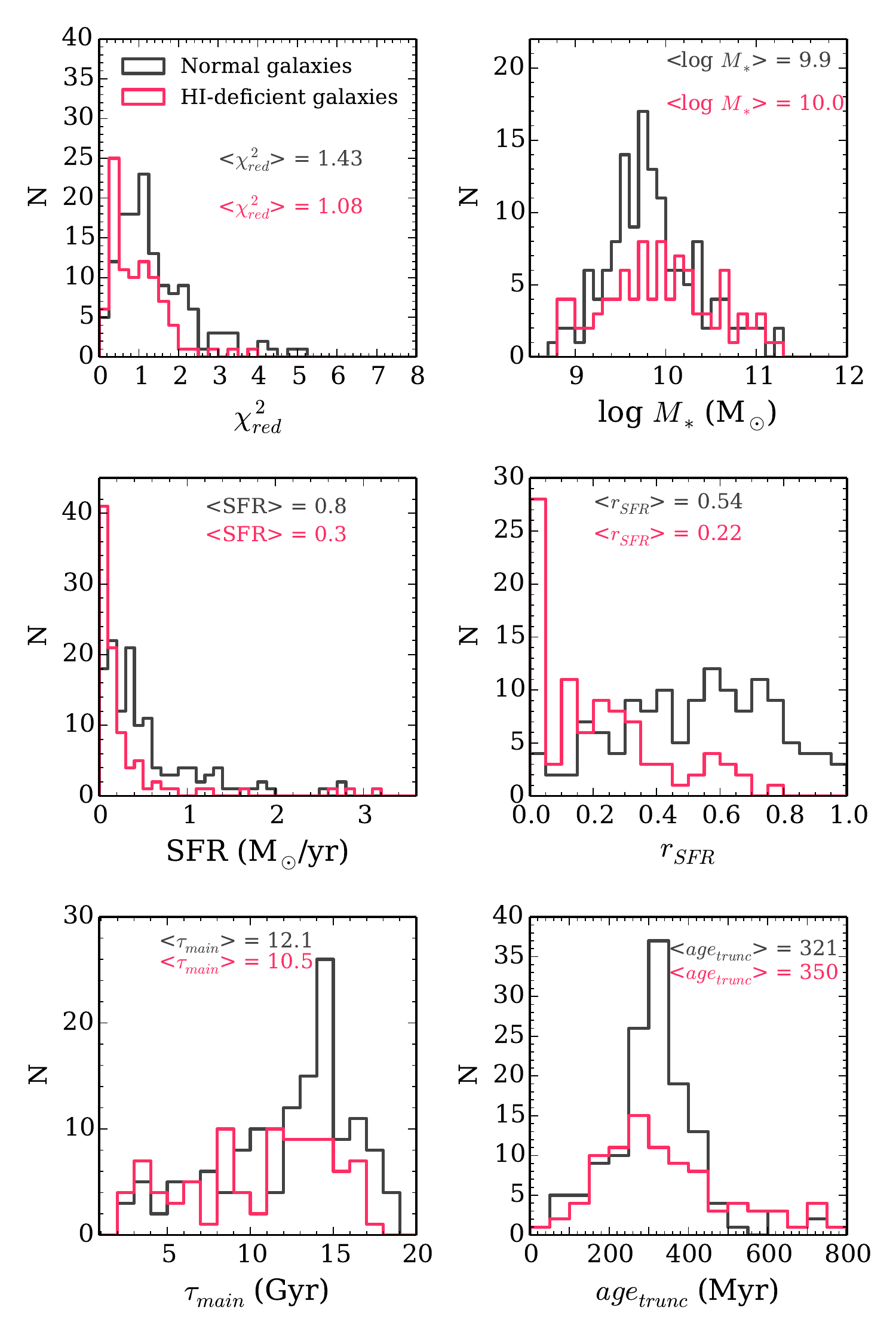}
  	\caption{ \label{hist_agetruncfree} Distribution of the output parameters obtained from the SED fitting procedure with CIGALE. The results for the normal galaxy sample are shown in grey while the results for the \HI-deficient sample are shown in red. }
\end{figure}

The distribution of the $\chi^2$, $M_*$, SFR, $r_{SFR}$, $\tau_{main}$, and $age_{trunc}$  obtained from the SED fitting, with $age_{trunc}$ left as a free parameter, are presented in Fig.~\ref{hist_agetruncfree} for both normal and \HI-deficient galaxies.
The $\chi^2$ distribution for both the normal and the \HI-deficient samples is similar with a mean value of 1.43 and 1.08 respectively.
The distributions of the stellar mass of both subsamples are similar, peaking at the same value than the results obtained from the run fixing $age_{trunc}$.
The distribution of star formation rates of the \HI-deficient clearly shows that most of these galaxies have a very low, almost zero.
The distribution of the $r_{SFR}$ parameter shows two different behaviors for normal and deficient galaxies.
The \HI-deficient subsample distribution shows lower values with 42 sources with $r_{SFR}<0.15$ out of 92 deficients galaxies.
The $r_{SFR}$ distribution of normal galaxies is flat and shifted toward larger values although not as close to 1 as one would expect.
The distribution of $age_{trunc}$ is very similar for both samples with about the same mean value (321\,Myr for the normal galaxies and 350\,Myr for the deficient sample) and same distribution.
This behavior reinforces our conclusion about the poor constraint on this parameter.
Finally the distribution of $\tau_{main}$ shows that most of the sources have $\tau_{main}\geq5$\,Gyr in this configuration too.  

\section{\label{nuvr}NUV-r versus \HIdef\ relation for the HRS galaxies}
\begin{figure}%[!h] 
  	\includegraphics[width=\columnwidth]{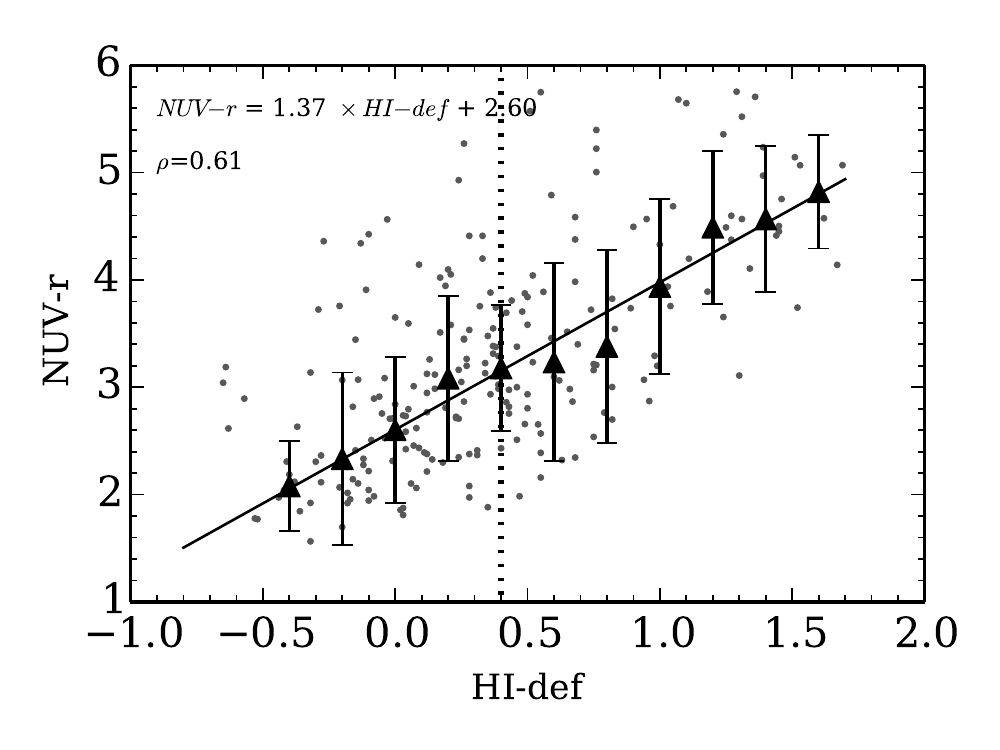}
  	\caption{ \label{nuvrdef}  NUV-r color as a function of \HIdef\ for the HRS galaxies studied in this work. Black triangles show the median values in 0.2 \HIdef\ bins, the error bars represent the standard deviation in each bin. The Spearman correlation coefficient and the results of the best linear fit are indicated. }
\end{figure}
It is established now that the NUV-r color is a good tracer of the gas content of galaxies \citep[e.g.,][]{Cortese11,Fabello11,Catinella13,Brown15}, we thus show in Fig.~\ref{nuvrdef} the NUV-r versus \HIdef\ relations for the galaxies studied in this work.
Using the same bins as in Fig.~\ref{rsfrvsdefhi}, we obtained a best linear fit of:

\begin{equation}
	NUV-r = 1.37 \times HI-def + 2.60,
\end{equation}
\noindent	associated with a Spearman correlation coefficient of $0.61$.

\end{document}